\documentclass[3p,times]{elsarticle}

\usepackage{ecrc}

\volume{00}

\firstpage{1}

\journalname{Nuclear Physics A}

\runauth{}

\jid{npa}

\jnltitlelogo{Nuclear Physics A}

\usepackage{amssymb}

\biboptions{square,comma,numbers,sort&compress}

\usepackage[figuresright]{rotating}

\begin{document}

\begin{frontmatter}



\dochead{}

\title{Azimuthal Jet Tomography at RHIC and LHC}


\author{Barbara Betz}
\author{Miklos Gyulassy}

\address{Institute for Theoretical Physics, Johann Wolfgang 
Goethe-University, 60438 Frankfurt am Main, Germany}
\address{Department of Physics, Columbia University, New York, 10027, USA}

\begin{abstract}
A generic jet-energy loss model that is coupled to state-of-the-art
hydrodynamic fields and interpolates between a wide class
of running coupling pQCD-based and AdS/CFT-inspired models is compared
to recent data on the azimuthal and transverse momentum dependence of
high-$pT$ pion nuclear modification factors and high-$pT$ elliptic flow
measured at RHIC and LHC. We find that RHIC data are surprisingly consistent
with various scenarios considered. However, extrapolations to LHC energies
favor running coupling pQCD-based models of jet-energy loss. While
conformal holographic models are shown to be inconsistent with data, recent non-conformal 
generalizations of AdS holography may provide an alternative description. 
\end{abstract}


\end{frontmatter}


\section{Introduction}
\label{Introduction}

Tomographic information about the evolution of the quark-gluon plasma (QGP)
formed in a high-energy nuclear collision is commonly extracted by jet-quenching
observables which, however, depend on the details of both the jet-medium dynamics,
$dE/dx\,(E,\vec{x},T)$, as well as the bulk temperature and flow velocity fields.
Recently, PHENIX \cite{Adare:2012wg} compared their measured data on the nuclear
modification factor in- and out-of-plane to different perturbative QCD (pQCD)-based 
and one AdS-CFT motivated jet-energy loss model, considering a longitudinal and
transverse expanding hydrodynamic background. Their result strongly suggest that
only the AdS-inspired model explains {\it both} the nuclear modification factor
and the high-$p_T$ elliptic flow accessible through 
$R_{\rm AA}^{\rm in/out}=R_{\rm AA}\left(1\pm2v_2\right)$. While 
the yield of nuclear modification factor could as well be described by the 
pQCD-based AMY, HT, and ASW models considered, their high-$p_T$ elliptic 
flow is significantly too low. This
is in line with the results in Ref.\ \cite{Molnar:2013eqa} and \cite{Jiechen2014}, 
showing that the pQCD (D)GLV energy loss prescription coupled to
either the 3D MPC parton transport model \cite{oai:arXiv.org:nucl-th/0005051} 
or the (2+1)D transverse expansion of VISH2+1 \cite{VISH} applying a temperature-independent running coupling
lead to a high-$p_T$ elliptic flow that is by a factor of $\sim 2$ smaller than the 
measured data, while the $R_{AA}$ can well be reproduced.
The aim of the present work is to braden the PHENIX analysis by considering a wider 
class of $dE/dx$ models coupled to different flow fields and by extending the 
analysis to a simultaneous description of RHIC and LHC. We study models based on 
pQCD, conformal and non-conformal AdS holography, as well as a
phenomenological prescription with an enhanced jet-energy loss around
the transition temperature of $T_c\approx 170$~MeV (referred to as SLTc \cite{Liao:2008dk}).
Each model is coupled to different bulk QGP collective fields as given by (1) VISH2+1 
\cite{VISH}, (2) the RL (Romatschke-Luzum) Hydro \cite{Luzum:2008cw}, 
and (3) a simple $v_{\perp}=0.6$ transverse blast wave model \cite{GVWH}.

The generic energy-loss model considered \cite{WHDG11,Betz:2012qq} is characterized by
the jet energy, path length, and thermal-field dependence described via the
three exponents $(a,b,c)$:
\begin{eqnarray}
\hspace*{-3ex}
\frac{dE}{dx}=\frac{dP}{d\tau}(\vec{x}_0,\phi,\tau)= 
-C_r\kappa(T)  P^a(\tau) \, \tau^{b} \, T^c
\;.
\label{Eq1}
\end{eqnarray}
Here, $T=T\,[\vec{x}(\tau)=\vec{x}_0+ (\tau-\tau_0) \hat{n}(\phi)]$
is the local temperature along the jet path at time $\tau$ for a jet
produced initially at time $\tau_0$ and distributed according to
a Glauber transverse initial profile \cite{Betz:2012qq}. For dimensionless couplings,
$c=2+z-a$. $\kappa(T)$ can depend on the local temperature field (SLTc model).
$C_r=1(\frac{9}{4})$ describes quark (gluon) jets. 

In the following, we particularly distinguish four different cases: (1) 
a running coupling QCD energy loss QCDrad $(a=0,b=1,c=3)$, (2) 
an elastic jet-energy loss \cite{WHDG11} QCDel $(a=0,b=0,c=2)$, (3) 
a scenario characterized by a conformal falling string \cite{AdS}
AdS $(a=0,b=2,c=4)$, and (4) a $T_c$-dominated energy loss 
\cite{Liao:2008dk} with $\kappa(T_c) =3 \kappa(\infty)$
SLTc $(a=0,b=1,c=3)$. For each scenario, the jet-medium coupling is adjusted to fit a 
single reference point at $p_T=7.5$~GeV for RHIC energies.

\section{Discussion and Results}
\label{Discussion}

\begin{figure}[t]
\center \includegraphics[scale=0.39]{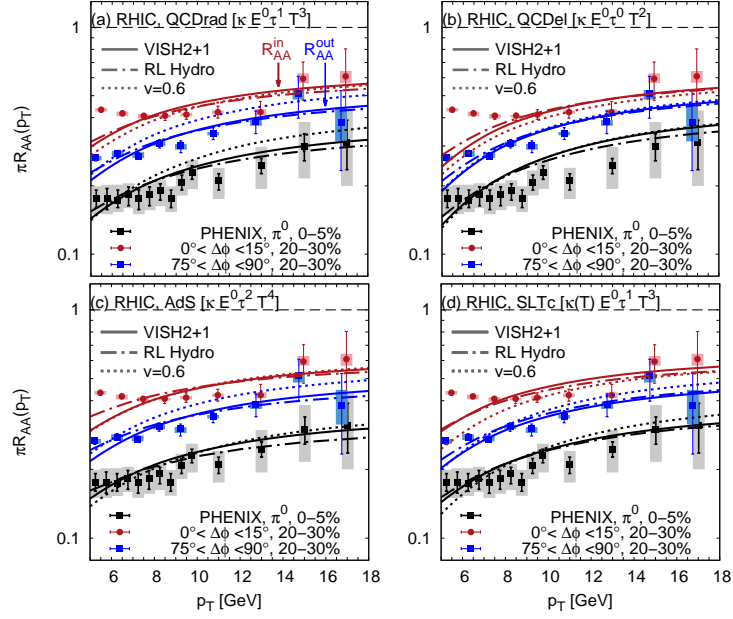}
\caption{In- and out-of-plane nuclear modification factors for 0-5\% and 20-30\% 
centrality at RHIC \protect{\cite{Adare:2012wg}}, 
compared to predictions simulating (a) a running coupling QCD energy loss,
(b) an elastic jet-energy loss, (c) a conformal AdS falling string scenario, and (d) 
the transition-phase dominated SLTc model. For each model, three different bulk evolutions are studied: 
ideal VISH2+1 \protect{\cite{VISH}} (solid), 
viscous $\eta/s=0.08$ RL Hydro \protect{\cite{Luzum:2008cw}} (dashed-dotted), 
and the $v_\perp=0.6$ blast wave model \protect{\cite{GVWH}} (dotted).}
\label{fig1}
\end{figure}

The most striking result in Fig.\ \ref{fig1} is that in contrast to the 
PHENIX results \cite{Adare:2012wg} {\it all} scenarios based on a (viscous)
hydrodynamic background describe the measured $p_T>8$~GeV data within
the present error bars. Only the $v_\perp=0.6$ transverse blast wave background 
leads to an in/out asymmetry with a factor of two
below recent PHENIX data. The qualitative difference to the PHENIX results,
in particular to the different slope of the $R_{\rm AA}^{\rm in/out}$,
is due to various details of the hydrodynamic simulation and the jet-energy
loss precription. In the study shown by PHENIX, the flow field was computed with an ideal 
(non-dissipative) hydrodynamic code assuming a Bag model first-order phase 
transition while the VISH2+1 code considered here is based on a smoothed (SM-EOS Q) 
equation of state (EoS) and the viscous RL Hydro employs a more realistic continuous
crossover transition.

\begin{figure}[t]
\begin{minipage}[t]{8.5cm}
\hspace*{-1cm}
\includegraphics[width=3.5in]{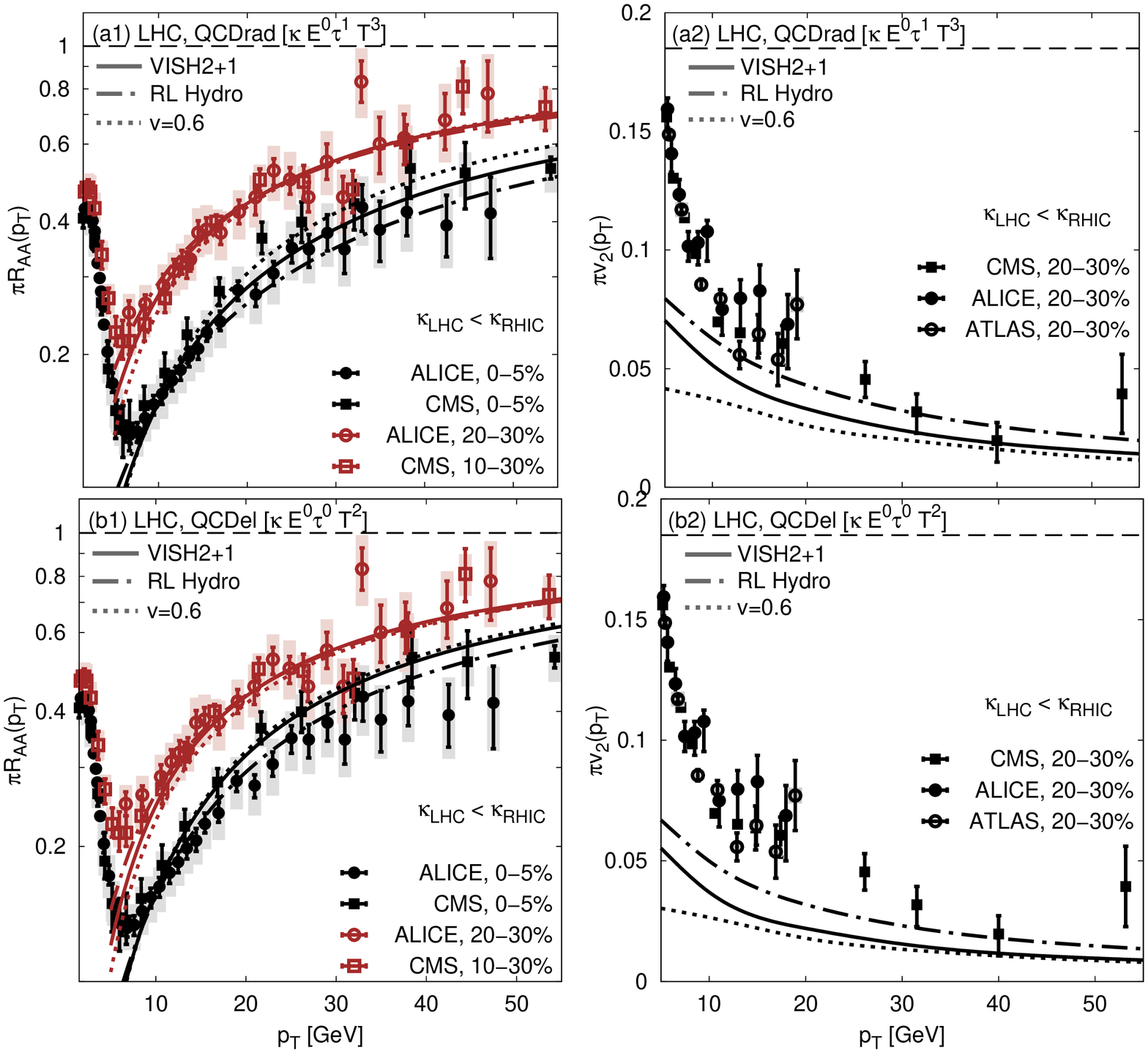}
\end{minipage}
\hspace*{-0.4cm}
\begin{minipage}[t]{8.5cm}
\hspace*{-0.5cm}
\includegraphics[width=3.5in]{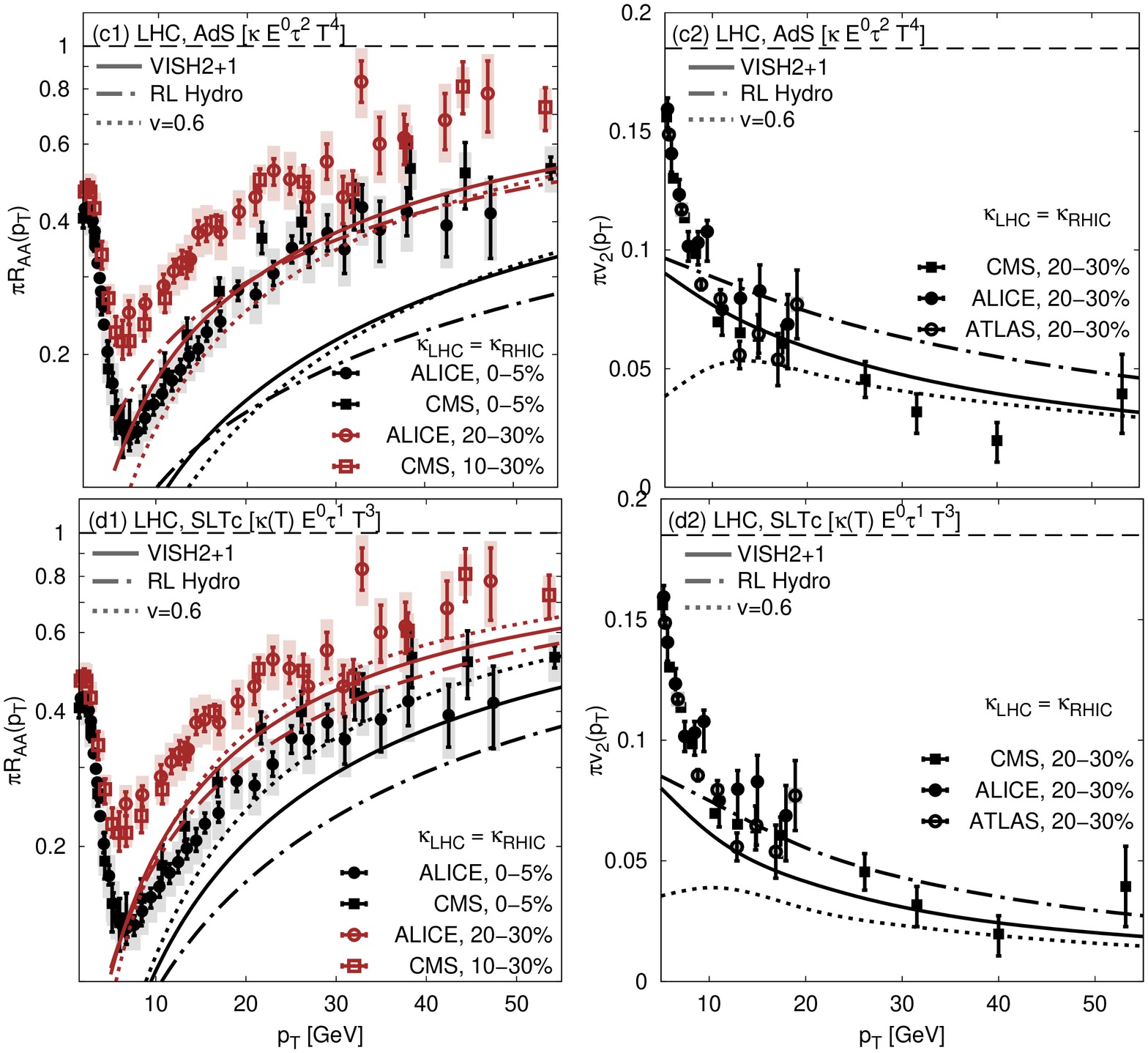}
\end{minipage}
\caption{The nuclear modification factors and high-$p_T$ elliptic flow for most central
and mid-peripheral centralities at LHC energies. The nuclear modification factor is taken from 
ALICE \protect{\cite{Abelev:2012hxa}} and CMS \protect{\cite{CMS:2012aa}}, the
high-$p_T$ elliptic flow is extracted from ALICE \protect{\cite{Abelev:2012di}}, 
CMS \protect{\cite{Chatrchyan:2012xq}}, and ATLAS \protect{\cite{ATLAS:2011ah}}.
The model calculations are the same as in Fig.\ \ref{fig1}. However,
the bulk QGP fields are taken from viscous ($\eta/s=0.08$) VISH2+1 \protect{\cite{VISH}} (solid), 
viscous $\eta/s=0.08$ RL Hydro \protect{\cite{Luzum:2008cw}} (dashed-dotted), 
and the $v_\perp=0.6$ blast wave model \protect{\cite{GVWH}} (dotted). For panels
(a) and (b), the jet-medium coupling $\kappa_{LHC}$ is reduced relative to RHIC to simulate a
QCD running coupling, while $\kappa_{LHC}=\kappa_{RHIC}$ is considered in panels (c) and (d), see text.}
\label{fig2}
\end{figure}

Given this difficulty of untangling the background and jet-energy loss effects
at one particular $\sqrt{s}$, we continue by exploring the dependence of the jet asymmetry
on the collision energy and study LHC conditions, see Fig.\ \ref{fig2}, for the
four jet-energy loss scenarios and the three background fields introduced above. However,
in contrast to RHIC data, at LHC energies the nuclear modification factors and the high-$p_T$ elliptic
flow are only issued separately (i.e.\ not as $R_{\rm AA}^{\rm in/out}$). 
Therefore, we have to split up the LHC results into two subplots for each
scenario (see Fig.\ \ref{fig2}).

For the pQCD-based scenarios QCDrad and QCDel, we assume a running coupling effect
\cite{WHDG11}. I.e.\ the jet-medium coupling $\kappa_{LHC}$ is reduced by $\sim (30-40)\%$ 
as compared to $\kappa_{RHIC}$. As shown in Fig.\ \ref{fig2}(a1) and (a2), QCDrad reproduces both
the nuclear modification factors and the high-$p_T$ elliptic flow within the
uncertainties of the bulk space-time evolution given, amongst others, by the initial condition, 
the shear viscosity over entropy ratio ($\eta/s$), and the initial time $\tau_0$. This
running coupling QCD energy loss, however, appears to be preferred
over the elastic jet-energy loss QCDel [Fig.\ \ref{fig2}(b1) and (b2)] .

In contrast to those pQCD cases discussed in Fig.\ \ref{fig2}(a) and (b), 
we do {\it not} consider a QCD running coupling effect for the AdS-inspired
energy loss and the SLTc model shown in Fig.\ \ref{fig2}(c) and (d) but assume that
$\kappa_{\rm LHC}=\kappa_{\rm RHIC}$. In the AdS case,
the reason is that {\it conformal} AdS/CFT implies scale invariance and thus 
does not allow for a running but only a fixed coupling. As can be seen from Fig.\ \ref{fig2}(c1),
this fixing of the jet-medium coupling at RHIC energies leads to the well-known
overquenching of the nuclear modification factor \cite{WHDG11} and thus to an
enhancement of the high-$p_T$ elliptic flow as compared to the pQCD scnearios.

This Fig.\ \ref{fig2}(c1) proofs that {\it conformal} AdS/CFT is ruled
out by the yield and the rapid rise of the $R_{\rm AA}(p_T)$ at LHC energies.

Since the SLTc model shown in Fig.\ \ref{fig2}(d1) and (d2) is a pQCD-based model as well,
a running coupling effect might apply. However, we choose the same jet-medium
coupling as for RHIC energies since it was shown in Ref.\ \cite{Liao:2008dk} that
the SLTc model can describe the nuclear modification factor and the high-$p_T$ 
elliptic flow both at RHIC and at LHC without any additional running coupling effect
for a purely longitudinally expanding medium. Thus, Fig.\ \ref{fig2}(d1) and (d2) proofs
that for a more realistic (2+1)D expanding medium the temperature dependence of the jet-medium
coupling around a critical temperature of $T_c=170$~MeV is not sufficient to describe
the nuclear modification factor at LHC energies.

\begin{figure}[t]
\center \includegraphics[scale=0.43]{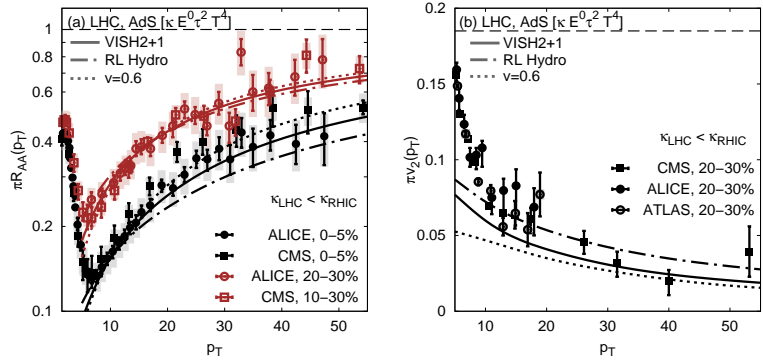}
\caption{The nuclear modification factors and high-$p_T$ elliptic flow for most central
and mid-peripheral centralities at LHC energies, cf.\ Fig.\ \ref{fig2},
compared to a non-conformal AdS-inspired energy-loss scenario with $dE/dx=\kappa E^0\tau^2T^4$,
allowing for a running coupling $\kappa_{LHC}<\kappa_{RHIC}$.}
\label{fig3}
\end{figure}

While we showed in Fig. \ref{fig2}(c1) and (c2) that {\it conformal} 
AdS/CFT is ruled out by the data, a recent non-standard AdS/CFT
formulation of ``shooting" strings \cite{ficnar13} leads to an analytic jet-energy loss
formula that interpolates between the above discussed extremes of QCDel and AdS (because the
path-length dependence within this prescription is itself temperature dependent). This ansatz suggests
that it might be necessary to broaden the scope of the AdS/CFT models 
to {\it non-conformal} scenarios. For a non-conformal AdS/CFT description, the
coupling constant can run. Allowing a running coupling effect for an
AdS/CFT-inspired energy loss of $dE/dx\sim E^0\tau^2 T^4$ results in 
nuclear modification factors and high-$p_T$ elliptic flow at LHC energies 
that are consistent with the measured data within the current experimental error bars as shown in Fig.\
\ref{fig3}.

\section{Summary}
\label{Summary}

We compare recent data on the nuclear modification factors
measured at RHIC \cite{Adare:2012wg} and LHC 
\cite{Abelev:2012hxa,Abelev:2012di,CMS:2012aa,Chatrchyan:2012xq,ATLAS:2011ah} 
to a wide class of jet-energy loss models \cite{Betz:2012qq} based on 
perturbative QCD (pQCD), conformal and non-conformal AdS holography, and a 
phenomenological model with an enhanced energy loss around
the transition temperature of $T_c\approx 170$~MeV,
coupled to different recent bulk QGP collective fields \cite{VISH,Luzum:2008cw,GVWH}. 
We found that (1) a running coupling pQCD energy loss seems to be favored,
(2) a realistic QGP flow backgrounds is essential
to simultaneously describe {\it both} the nuclear modification factors and 
the high-$p_T$ elliptic flow, and (3) {\it conformal} AdS string-like jet
holography appears to be ruled out by the LHC data while novel non-conformal
generaliztaions of string models \cite{ficnar13} may provide an alternative
description. Further details of the present study, especially a discussion on the
differences between the energy-loss model studied here and other pQCD energy-loss 
prescriptions the will be presented elsewhere \cite{Betz2014}.

\section{Acknowledgements}

We thank P.\ Romatschke, U.\ Heinz, and C.\ Shen for making their 
hydrodynamic field grids available to us and J.\ Xu for the direct
comparison and cross-check with the CUJET2.0 results. 
This work was supported in part through the Helmholtz International Centre 
for FAIR within the framework of the LOEWE program (Landesoffensive zur Ent\-wicklung
Wissenschaftlich-\"Okonomischer Exzellenz) launched by the State of
Hesse, the US-DOE Nuclear Science Grant No.\ DE-AC02-05CH11231 
within the framework of the JET Topical Collaboration, and the
US-DOE Nuclear Science Grant No.\ DE-FG02-93ER40764.

\end{document}